# A Comb-based Colorless Coherent WDM Transmitter

**D. Che[1*], B. Stern[1], K. Kim[1], C. Ozdilek[2], T. Shpakovsky[2], J. D. Jost[2], M. Karpov[2]**

[1] *Nokia Bell Labs, Murray Hill, NJ 07974, United States*

[2] *Enlightra, Rue de Lausanne 64, Renens, 1020, VD, Switzerland*

[*] *di.che@nokia-bell-labs.com*

**Abstract:** We propose a comb-based WDM transmitter capable of modulating independent signals to comb lines without demultiplexing them and prove its concept and potential scalability in a WDM transmitter consisting of a Kerr microcomb and a silicon I/Q modulator array.

## 1. Introduction

Chip-scale optical frequency combs (OFC) [1,2] have been actively developed over the years as a promising multi-channel light source for highly parallel wavelength-division multiplexing (WDM) transmissions. The co-integration of an OFC and parallel modulators onto one photonic integrated circuit (PIC) requires a wavelength demultiplexer in between, which separates the comb lines as single-wavelength ones to be fed to each modulator individually. While demultiplexers (or wavelength selective switches (WSS) in a more general case) have been commonly deployed in a module form and used in microcomb-based WDM demonstrations [1,2], implementing them on PICs is a challenging task especially when scaling to a higher channel count. The difficulty not only comes from the fabrication of a high-quality filter bank with low insertion loss and high extinction ratio among channels, but also the sophisticated on-chip wavelength control that can be power hungry. By cascading resonators along a single bus waveguide to selectively encode data onto wavelengths aligned to different resonances, a microresonator-based modulator array can avoid demultiplexers [3]. However, such a transmitter is still wavelength selective that requires dedicated and precise control of individual resonators, adding complexity and power consumption for tuning and stabilization.

Inspired by the optical time-division multiplexing (OTDM) concept [4], we recently demonstrated a comb-based WDM system [5] without using demultiplexers. The idea is to add a series of time delays to power-split OFC copies that introduce orthogonality among them to enable digital demultiplexing of the signals modulated on parallel channels by multi-input-multi-output (MIMO) equalization. However, the scheme requires simultaneous detection of all WDM channels to support MIMO co-processing in one receiver, thus limiting its application to point-to-point transmissions. For future chip-scale coherent transmitters with tens of WDM channels, it is highly desired to independently modulate and freely route each channel. Here, we propose an OFC-based coherent WDM transmitter to achieve an arbitrary signal generation on each comb line without demultiplexing the lines. Compared to a traditional demultiplexer-based transmitter, it does not need extra hardware, and the only added complexity is from a simple MIMO pre-equalization (PEQ) which consumes only a small portion of power in coherent DSP [6]. Distinguished from prior OTDM schemes [4], it needs neither transmitter-receiver synchronization/locking nor the locking of OFC free spectral range (FSR) to RF signals inside a transmitter, that unlocks a free-running OFC as the light source. We use a silicon-nitride ($Si_3N_4$) microcomb [1] and a silicon-photonic (SiPh) modulator array to prove the concept for high-density integration.

## 2. Principle

We assume an $N$-channel transmitter in Fig. 1, using an OFC with $N$ lines followed by an array of $N$ I/Q modulators (IQM). Without a demultiplexer, each modulator generates $N$ identical replicas of its input signal evenly spaced on the spectrum. Each wavelength mixes all the $N$ signals after colorless power combining all branches. A series of time

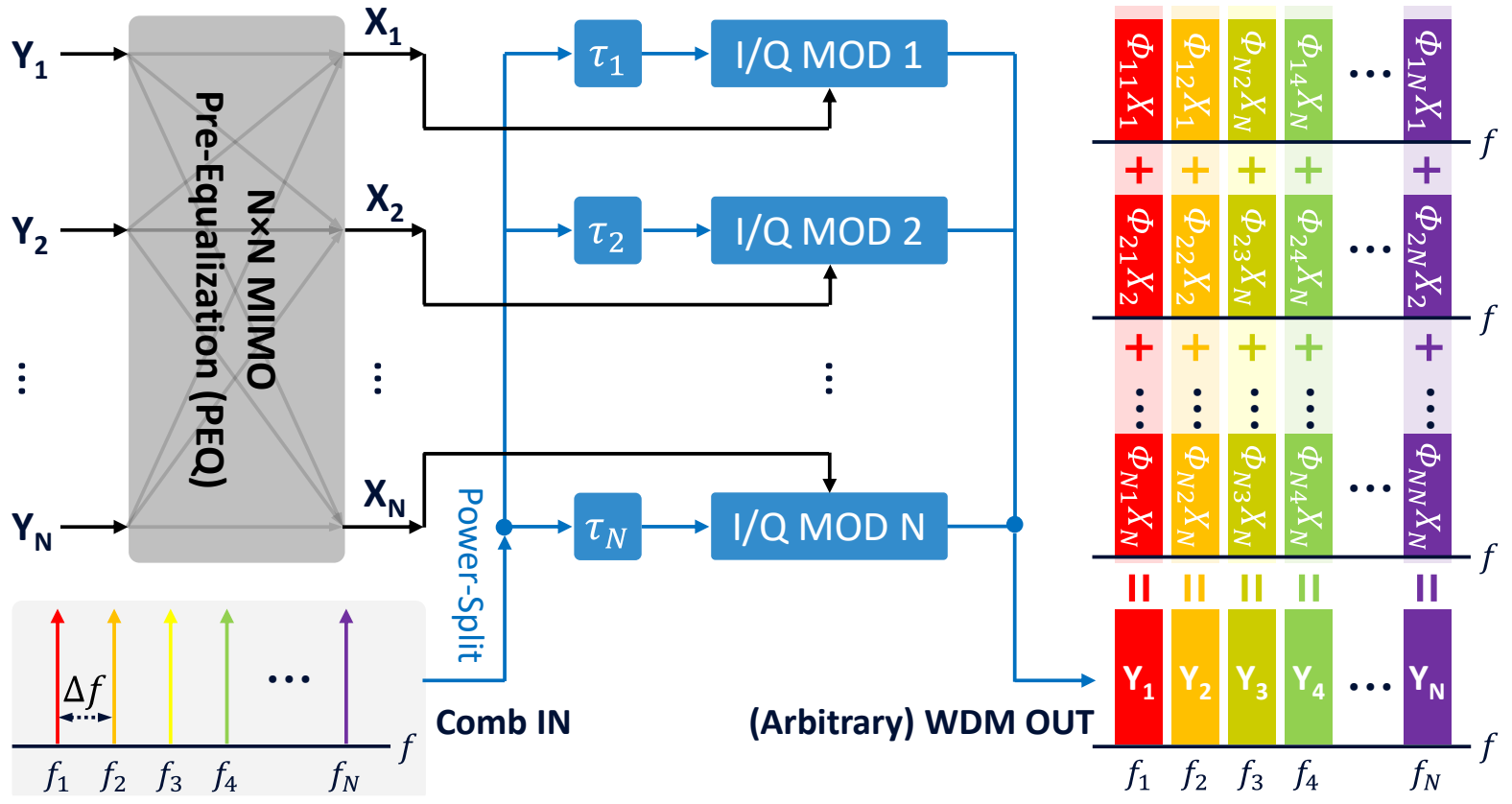

$$\vec{\varphi}_n = 2\pi\Delta f\cdot[0\quad \tau_n\quad 2\tau_n\quad 3\tau_n\quad \cdots\quad (N-1)\tau_n] \quad \textbf{(1)}$$

$$\boldsymbol{\varphi} = 2\pi\Delta f\cdot\begin{bmatrix}0 & 0 & 0 & 0 & \cdots & 0\\ 0 & \tau_2 & 2\tau_2 & 3\tau_2 & \cdots & (N-1)\tau_2\\ \vdots & \vdots & \vdots & \vdots & \ddots & \vdots\\ 0 & \tau_N & 2\tau_N & 3\tau_N & \cdots & (N-1)\tau_N\end{bmatrix}$$

$$denote:\ \boldsymbol{\Phi} = \exp j\boldsymbol{\varphi} \quad \textbf{(2)}$$

$$\tau_n = \frac{(n-1)}{N}\cdot\frac{1}{\Delta f} \quad \textbf{(3)}$$

$$\begin{bmatrix}Y_1\\ Y_2\\ \vdots\\ Y_N\end{bmatrix} = \begin{bmatrix}A_1\Phi_{11} & A_1\Phi_{21} & \cdots & A_1\Phi_{N1}\\ A_2\Phi_{12} & A_2\Phi_{22} & \cdots & A_2\Phi_{N2}\\ \vdots & \vdots & \ddots & \vdots\\ A_N\Phi_{1N} & A_N\Phi_{2N} & \cdots & A_N\Phi_{NN}\end{bmatrix}\begin{bmatrix}X_1\\ X_2\\ \vdots\\ X_N\end{bmatrix}$$

$$\vec{\boldsymbol{Y}} = (\boldsymbol{\Phi}\mathbf{A})^T\cdot\vec{\boldsymbol{X}} \quad \textbf{(4.1)}$$

$$\vec{\boldsymbol{X}} = \left((\boldsymbol{\Phi}\mathbf{A})^T\right)^{-1}\cdot\vec{\boldsymbol{Y}} \quad \textbf{(4.2)}$$

Fig. 1. A demultiplexer-free $N$-channel WDM transmitter. $\boldsymbol{\Phi}$: phase matrix (phase shifts of $N$ comb lines on $N$ branches); $\vec{\boldsymbol{Y}}$: target WDM signals; $\vec{\boldsymbol{X}}$: IQM drive signals; $\boldsymbol{A}$: a diagonal matrix with comb line amplitudes $(A_1, A_2, \ldots A_n)$ on the diagonal.

delays are added to the branches. According to the time-shifting property of Fourier transform, a time delay results in phase variations of frequency components proportional to the frequency. Hence, the phase variations of the comb lines for the $n$-th branch ($\vec{\varphi}_n$) due to the delay $\tau_n$ form an arithmetic sequence spacing at $2\pi\Delta f \cdot \tau_n$ ($\Delta f$ is the comb FSR) as shown in Eq. (1), which takes the 1st comb line as a reference with zero phase variation. We define a matrix $\boldsymbol{\varphi}$ that describes the phase variations of the $N$ comb lines for all the $N$ branches as Eq. (2) and define $\boldsymbol{\Phi} = \exp j\boldsymbol{\varphi}$. As proven in [5], if the delays evenly segment the OFC repetition period ($T = 1/\Delta f$) as in Eq. (3), $\boldsymbol{\Phi}$ would be unitary. Denoting the input signals to the modulator array as $\vec{\boldsymbol{X}} = [X_1, X_2, \dots, X_N]^T$ and the WDM signal at the transmitter output as $\vec{\boldsymbol{Y}} = [Y_1, Y_2, \dots, Y_N]^T$, the signal at each wavelength ($Y_n$) is a linear combination of a phase-shifted series of $\vec{\boldsymbol{X}}$, namely, $\vec{\boldsymbol{Y}} = \boldsymbol{\Phi}^T\vec{\boldsymbol{X}}$ (Eq. (4.1)). The equations can also include amplitude terms ($\boldsymbol{A}$) to accommodate a non-flat OFC power profile.

The unitary nature of $\boldsymbol{\Phi}$ guarantees an information-lossless transformation between $\vec{\boldsymbol{Y}}$ and $\vec{\boldsymbol{X}}$. To generate a target WDM signal of $\vec{\boldsymbol{Y}}$, we can perform a MIMO-PEQ of $\vec{\boldsymbol{X}} = (\boldsymbol{\Phi}^T)^{-1}\vec{\boldsymbol{Y}}$ (Eq. (4.2)) and then apply $\vec{\boldsymbol{X}}$ as the drive signals of IQMs. Simply put, a demultiplexer-free transmitter taking $\vec{\boldsymbol{X}}$ as an input is equivalent to a demultiplexer-based transmitter (where each comb line is modulated individually) taking $\vec{\boldsymbol{Y}}$ as an input. Since $\boldsymbol{\Phi}$ is only determined by the on-chip delays and comb FSR, the MIMO-PEQ is independent of RF signals $\vec{\boldsymbol{X}}$. The scheme supports any modulation formats and symbol rates, so long as the signal bandwidth per channel does not exceed $\Delta f$. The MIMO-PEQ can be a single-tap filter, which requires very small pre-processing complexity even when scaling to tens of WDM channels.

## 3. Proof-of-concept of MIMO-PEQ using a SiPh modulator array with 4×633-Gb/s net data rate

We fabricated a SiPh chip (Fig. 2a) with an array of 4 IQMs on 220-nm SOI wafers. Each IQM consists of 2 Mach-Zehnder modulators (MZM) [7] and 3 thermal phase shifters for bias control. The measured 3-dB E/O bandwidth is about 34 GHz. The input is equally split into 4 ways using a broadband multimode interference (MMI) coupler, and the IQM outputs are recombined with a 2nd MMI. The 4 branches have varying waveguide lengths to provide relative delays of 0, 3.06, 6.12, and 9.18 ps, respectively, leading to an optimal FSR of 81.7 GHz for a unitary $\boldsymbol{\Phi}$ matrix based on Eq. (3). The delay deviation due to fabrication in waveguide width is estimated to be lower than 10 fs based on typical foundry cross-wafer statistics [7]. This can satisfy the required precision even with tens of on-chip modulators.

16-QAM or probabilistically-shaped (PS) 64-QAM signals were digitally generated and pulse-shaped by a root-raised cosine filter (RRC) with 0.01 roll-off. The 4 pairs of I/Q signals were pre-processed (Eq. (4.2)) with a *single-tap* 4×4 MIMO filter. We used a coherent receiver to calibrate the phase term in the $\boldsymbol{\Phi}$ matrix. For each column of $\boldsymbol{\Phi}$ (corresponding to each comb line), we aligned the local oscillator (LO) frequency to that comb line and measured the optical phase for each IQM at the chip output. The amplitude term can also be easily calibrated and included in Eq. (4.2), but we kept the comb flat in power in this preliminary work. A 128-GSa/s arbitrary waveform generator (AWG) with 8 synchronized channels generates the 4 I/Q signals. Due to the high $V_\pi$ (estimated to be ~8.3V) of the MZM, each AWG output was amplified with a 23-dB gain RF driver to provide a peak-to-peak swing of about 5V. We used an E/O comb to test the MIMO-PEQ DSP, which was generated by driving a phase modulator (PM) with a 40.4-GHz RF followed by a WSS which selected alternating lines to get an FSR of 80.8 GHz and 4 lines in total. With an 80-GBd signal (0.01 roll-off) per channel, the 4-channel WDM signal formed a 323.2-GHz superchannel (Fig. 2c). The signal then went through a polarization-division multiplexing (PDM) emulator and was detected by a dual-polarization coherent receiver, which used an integrated tunable laser assembly (ITLA) as an LO. The ITLA tuned its frequency to each WDM channel and detected them one at a time. We assume a rate-0.8262 soft-decision (SD) forward error correction (FEC) code with a normalized generalized mutual information (NGMI) threshold of 0.8714 [8]. The maximum entropy that satisfies this threshold is 5 bits/symbol, leading to a total net data rate of 4×633 Gb/s (Fig. 2d).

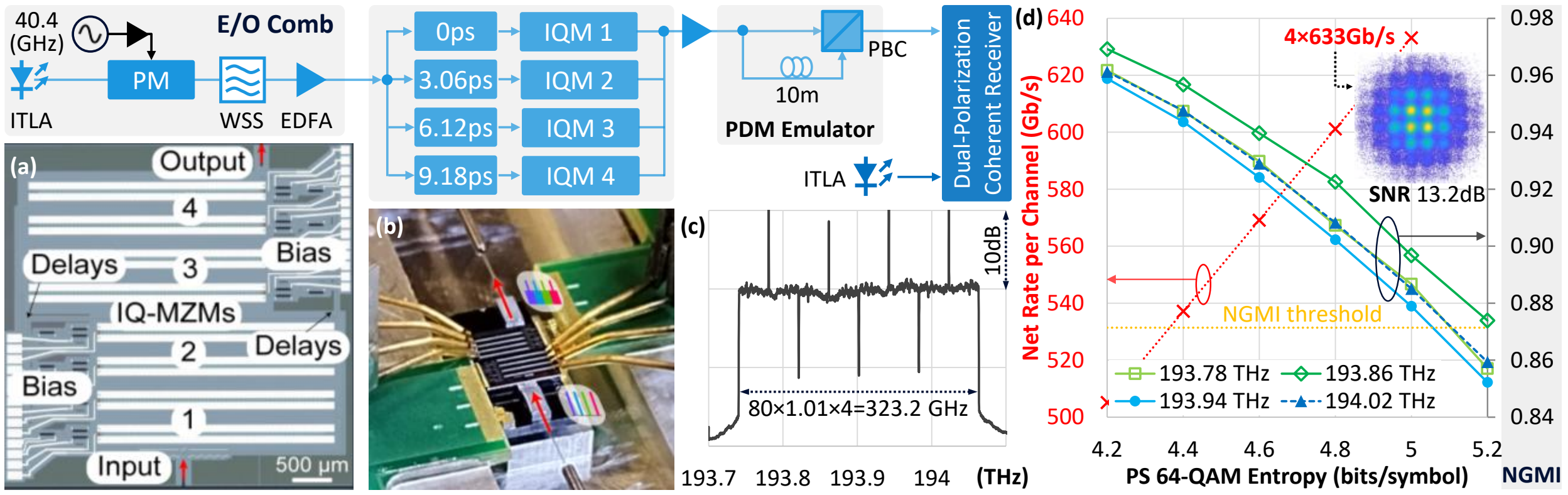

Fig. 2. The SiPh modulator array and MIMO-PEQ DSP validation setup. (a) The layout and (b) a photo of the chip, with phase shifter connections wirebonded to a PCB, and in/out fibers attached. The IQMs are interfaced with multi-contact RF probes (on both sides of the chip). (c) 4-channel WDM spectrum, each channel carrying a PDM 80-GBd PS 64-QAM signal. (d) NGMI of the 4 channels as a function of PS-QAM entropies.

## 4. Proof of feasibility of combining a $Si_3N_4$ microcomb and the SiPh modulator array

We use a dissipative Kerr soliton (DKS) microcomb to verify the feasibility of using a free-running OFC (*i.e.*, no need of an RF reference) with the SiPh modulator array. The combination of the wideband microresonator-generated OFC and the SiPh modulators reveal the potential scalability to a higher WDM channel count. The OFC consists of 3 main components: a continuous wave pump laser, a $Si_3N_4$ microresonator (Q-factor >1 million), and control electronics for managing the internal components and OFC parameters. The devices are packaged into a benchtop module. The comb is centered at 193.39 THz, with an FSR of about 100 GHz, a >2-THz 3-dB bandwidth and a >40-dB optical signal-to-noise ratio (OSNR). Its integrated feedback loop ensures long-term stability of the soliton state. As the delays on the SiPh chip were designed for ~80-GHz FSR, the transfer matrix ($\boldsymbol{\Phi}$) would not be well-conditioned if the input was 4 comb lines spacing at 100 GHz. However, if we use 3 IQM branches with delays of 0, 3.06, and 6.12 ps (leading to an optimal FSR of 108.9 GHz), the 3×3 $\boldsymbol{\Phi}$ matrix would be close to unitary. Thus, we reduce the WDM channel count to 3 in this section, without the loss of effectiveness of verifying the MIMO-PEQ. We fixed the modulation as single-polarization 16-QAM and use SNR as the metric to provide a direct comparison among various system conditions.

We first changed the symbol rates from 70 to 100 GBd to prove the RF signals can be fully independent from the OFC. As shown in Fig. 3a, the system works well for all the symbol rates, except for some SNR penalties with higher symbol rates due to the bandwidth limit of MZMs. To verify the colorless nature of the scheme, we chose 5 wavelength windows across the C-band from 191.5 to 194.8 THz (Fig. 3b). By re-calibrating and applying a different $\boldsymbol{\Phi}$ matrix to the I/Q signals for each window, all the wavelengths (with 80-GBd signals) report stable SNR around 13 dB (Fig. 3c). Different from a demultiplexer based transmitter, the proposed technique may put a stricter requirement on microcomb stability, because the suppression of inter-channel crosstalk relies on a precise MIMO-PEQ filter, whose coefficients would change if the relative power or phase among the comb lines fluctuates. We performed a reliability test over 26 hours by keeping all the optics on and monitoring the SNR of a channel centered at 193.89 THz (with 3 WDM channels turned on). The SNR remains stable (Fig. 3d), indicating the MIMO-PEQ can provide reliable crosstalk cancellation.

## 5. Conclusion

We propose a new chip-scale solution for highly-parallel coherent transmitters. It simplifies the monolithic integration of an OFC and a modulator array with any integration platform like silicon or thin-film lithium niobate [9]. The color-less nature allows for scaling to a larger number of integrated modulators without requiring individual wavelength control that relieves the complexity and power consumption. Together with on-chip amplification (*e.g.*, [10]), it would be feasible to develop a DWDM coherent transmitter with extremely high data-rate density in an order of Tb/s/mm$^2$.

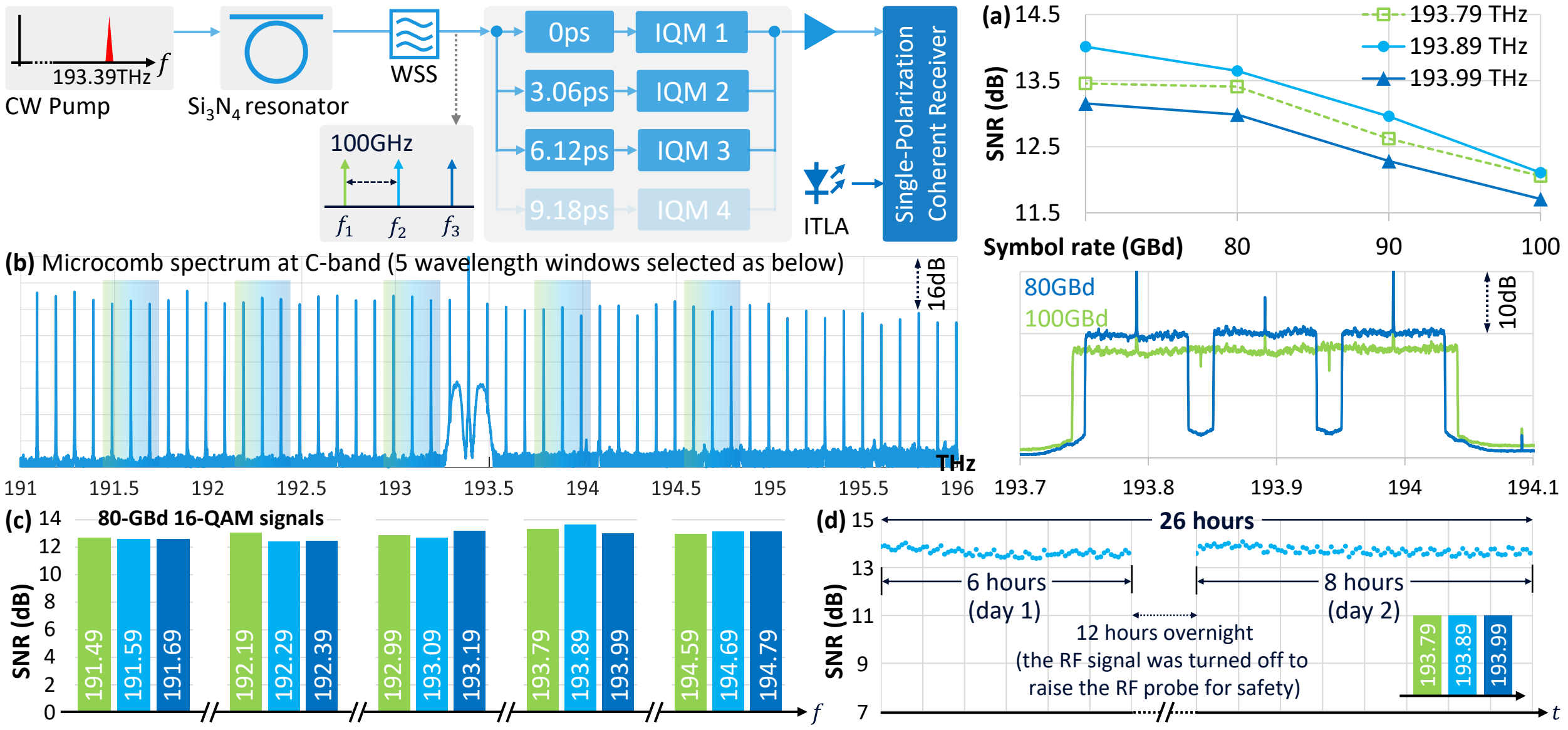

Fig. 3. A demultiplexer-free 3-channel WDM transmitter using a DKS microcomb (100-GHz FSR). (a) System SNR as a function of symbol rates. (b) Microcomb spectrum at C-band. (c) Demonstration of the colorless nature of the scheme: the system SNR are similar (at around 13 dB) with 5 wavelength windows across the C-band. (d) Reliability test for the MIMO-PEQ by monitoring the SNR of an 80-GBd signal at 193.89 THz.